\documentclass{svmult}

\usepackage{type1cm}        
\usepackage{makeidx}         
\usepackage{graphicx}        
\usepackage{multicol}        
\usepackage[bottom]{footmisc}

\usepackage{newtxtext}       %
\usepackage[varvw]{newtxmath}       

\usepackage{natbib}
\setcitestyle{round,aysep={},yysep={;}}

\makeindex             

\begin{document}

\title*{Unraveling RNA by Mechanical Unzipping}
\author{Paolo Rissone \and Isabel Pastor \and Felix Ritort}
\institute{Paolo Rissone 
\and
Isabel Pastor
\and
Felix Ritort
\at Facultat de Fisica, Universitat de Barcelona, Carrer de Martí i Franqués 1, Barcelona 08028, Spain
\and Felix Ritort
\at
Institut de Nanoci\`encia i Nanotecnologia (IN2UB), Facultat de Fisica, Universitat de Barcelona, Carrer de Martí i Franqués 1, Barcelona 08028, Spain, \email{fritort@gmail.com}}

%
%

\maketitle

\abstract*{We review the basic concepts and tools for mechanically unzipping RNA hairpins using force spectroscopy. By pulling apart the ends of an RNA molecule using optical tweezers, it is possible to measure the folding free energy at varying experimental conditions. Energy measurements permit us to characterize the thermodynamics of RNA hybridization (base pairing and stacking), the dynamics of the formation of native and kinetic (intermediates and misfolded) molecular states, and interactions with metallic ions. This paper introduces basic concepts and reviews recent developments related to RNA force thermodynamics, native and barrier RNA energy landscapes, and RNA folding dynamics. We emphasize the implications of mechanical unzipping experiments to understand non-coding RNAs and RNAs in extreme environments.
}

\abstract{We review the basic concepts and tools for mechanically unzipping RNA hairpins using force spectroscopy. By pulling apart the ends of an RNA molecule using optical tweezers, it is possible to measure the folding free energy at varying experimental conditions. Energy measurements permit us to characterize the thermodynamics of RNA hybridization (base pairing and stacking), the dynamics of the formation of native and kinetic (intermediates and misfolded) molecular states, and interactions with metallic ions. This paper introduces basic concepts and reviews recent developments related to RNA force thermodynamics, native and barrier RNA energy landscapes, and RNA folding dynamics. We emphasize the implications of mechanical unzipping experiments to understand non-coding RNAs and RNAs in extreme environments.
}

\section{Introduction}
\label{sec:intro}
RNA, the genome's dark matter, directly impacts biological diversity and life \citep{darnell2011rna,filipowicz2022rna}. RNAs can fold into multiple configurations stabilized by secondary and tertiary structures \citep{butcher2011molecular,herschlag2018story}, multivalent cations, and ligands \citep{pyle2002metal,woodson2005metal,draper2005ions, bowman2012cations}. The promiscuity of base pairing and stacking interactions makes RNA a unique biopolymer with many functions, from information carrier to regulatory and enzymatic activity. RNA exhibits a significant degree of heterogeneity at the sequence level and the conformational level \citep{treiber2001beyond,russell2002rapid, brion1997hierarchy, cruz2009dynamic}. Upon folding, RNA can form native and non-native structures (such as misfolded and intermediates) \citep{woodson2010compact}, with critical roles at the level of genomic maintenance and the cellular function \citep{mattick2006non,aalto2012small}, therapeutics \citep{esteller2011non,matsui2017non} and diseases \citep{jain2017rna, blaszczyk2017structures, zhao2021dys}. Although the molecular forces operating in RNA are known, the role played by disorder at the structural and functional levels poses severe challenges to the life scientist who must cope with unprecedented complexity.

In recent years, a knowledge gap has appeared not only at the level of RNA transcriptomics but also at the level of non-coding RNAs (ncRNAs) and their remarkable variety of functions in concert with ligands and proteins \citep{stefani2008small, aalto2012small,mercer2009long,ma2013classification, fatica2014long, statello2021gene}. 
 
Although the role of many RNAs remains unknown, new RNAs with new functionalities and structures are being discovered. Besides the much-studied tRNA, rRNA, microRNA, riboswitches, ribozymes, and artificially evolved RNAs, novel behaviors have been observed in response to environmental cues such as temperature (e.g. RNA cold denaturation \citep{mikulecky2002cold}, RNA thermometers \citep{loh2013temperature}), and in concerted action with proteins (catalytic complexes, chaperones, packaging, condensation, etc.). 

Despite the enormous progress in next-generation sequencing and big data analysis, our current knowledge of RNA diversity is compromised by the limited accuracy, sensitivity, and specificity of available methods to detect different RNA conformations across RNA populations. Moreover, determining the folding pathways and the energetics of the various RNA structures is essential to understanding RNA function. Single-molecule techniques have represented a big step in addressing RNA complexity \citep{ritort2006single,seidel2007single}. Their great sensitivity and accuracy permit us to detect and measure the folding energies of rarely occurring conformations that escape detection by the standard bulk methods. Powerful techniques such as single-molecule FRET \citep{zhuang2005single,aleman2008exploring} and force spectroscopy \citep{ritchie2015probing,bustamante2021optical} can monitor RNA conformational transitions in real-time 
More recently, solid-state nanopore microscopy for RNA target detection can analyze thousands of single RNAs without amplification offering exciting prospects \citep{henley2016studies,bovskovic2022nanopore}. 

Compared to DNA, RNA exhibits more complex behavior. The replacement of deoxyribose for ribose and thymine for uracil makes RNA catalytic due to the reactive polarizable 2'-OH group of ribose. Ribose also induces large changes at the level of base stacking interactions between contiguous bases. In their double-stranded forms, nucleic acids (NA) form distinct right-handed double helices, B-form and A-form. Although DNA can adopt both A-form and B-form, RNA is mainly found in A-form with only a few rare exceptions \citep{shi2003structure}. In DNA, bases are mainly parallel to the helical plane with an interphosphate distance of 3.4\AA.
In contrast, RNA bases are tilted by approximately 19 degrees relative to the helical plane and the interphosphate distance is smaller ($\sim 2.8$\AA); These structural differences generate stacking between inter-strand bases and tighter water molecular bridges between phosphates and bases in RNA. Overall, base stacking tends to be stronger in RNA than in DNA. Base stacking is due to the Van der Waals attractive forces of the fluctuating dipole-dipole interactions between contiguous bases. Much weaker than the covalent nature of hydrogen bonding, the effect of the latter is minimized upon secondary structure formation due to the compensation effect of hydrogen bonding with water. Overall, base stacking and hydrogen bonding contribute equally to RNA helix stabilization, albeit the $1/r^6$ dependence of Van der Waals forces makes stacking strongly sensitive to the inter-base distance, $r$. Therefore, RNA structure strongly depends on RNA stacking between intra-strand and inter-strand bases, making RNA folding prediction a difficult problem.  

Here we briefly review recent discoveries in the thermodynamics of RNA folding using force spectroscopy studies with laser optical tweezers (LOT). The paper is organized as follows. Section \ref{sec:SM} describes three main experimental techniques to investigate RNA kinetics and thermodynamics at the single-molecule level. In Sec. \ref{sec:RNApulling} we focus on RNA force spectroscopy, the main experimental protocols and a few selected exemples. In Sec. \ref{sec:energetics} we discuss how RNA unzipping experiments make it possible to derive RNA thermodynamics at 0.1kcal/mol accuracy. Section \ref{sec:dynamics} introduces the concept of the barrier energy landscape and the importance of stem-loops to stabilize a multiplicity of RNA kinetic structures. Finally, in Sec. \ref{sec:future} we digress about future perspectives in single-RNA manipulation.

\section{The Power of Single RNA Manipulation}
\label{sec:SM}

\begin{figure}[h]
\centering
\includegraphics[width=\textwidth]{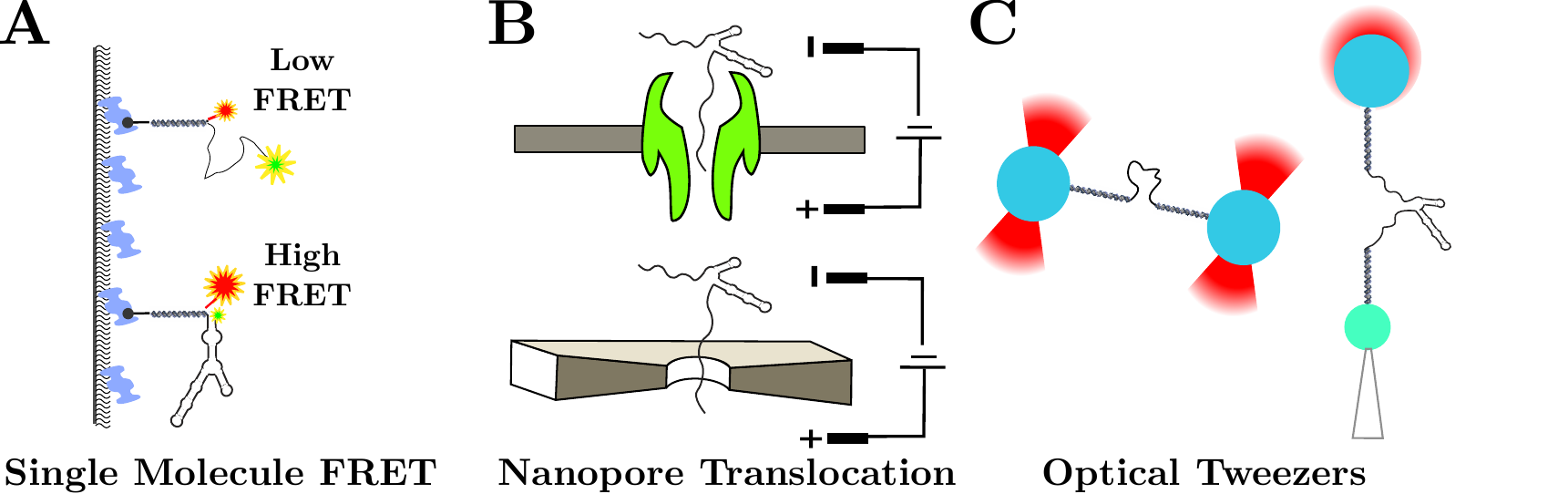}
\caption{\label{fig:Tech} Single-molecule techniques. (A) On-surface-immobilized smFRET experiment. (B) Nanopore microscopy uses a biological pore (top panel) and a solid-state pore (bottom panel). (C) Optical tweezers experiments in a double-trap geometry (left panel and in a single-trap geometry with a micropipette (right panel).   } 
\end{figure}

Single-molecule (SM) experiments permit us to study individual molecules' behavior and characterize the molecular properties' heterogeneity. On the contrary, in conventional bulk experiments, any observation is an average over a large population of molecules, making it impossible to observe ephemeral or transient events.
During the last decades, the development of single-molecule techniques opened a new window to the rich phenomenology of RNA molecules. In particular, three useful single-molecule techniques are single-molecule fluorescence (Förster) resonance energy transfer (smFRET) \citep{aleman2008exploring,zhuang2005single,ray2018life,chauvier2019probing, ha1999ligand, zhao2009rna}, solid-state nanopore microscopy \citep{seidel2007single,cui2021recent,bandarkar2020nanopore,lee2021tertiary,bovskovic2022nanopore} and force spectroscopy using optical tweezers \citep{woodside2006direct,bustamante1991entropic, ritchie2015probing, manosas2005thermodynamic, wen2007force, zhuang2005single}. First, we overview these techniques, focusing next on the optical tweezers experiments.

\subsection{smFRET Experiments}
\label{subsec:smFRET}
smFRET measurements measure conformational changes with a high temporal resolution (ms) \citep{joo2008advances,ha2001single}. This technique measures the distance between two fluorophores (donor and acceptor) at different positions along the biomolecule. The donor is located in the vicinity of the acceptor (distance less than 10 nm) and excited with an appropriate light wavelength. Part of the energy emitted by the donor is transferred to the acceptor through a non-radiative dipole-dipole interaction, causing the acceptor's emission. Donor-acceptor energy transfer is a quantum-mechanical effect due to the overlap between the donor emission spectrum and the acceptor absorption spectrum. The energy transfer efficiency, defined as the ratio $I_A/(I_D+\eta I_A)$ between light intensity emitted by the donor $I_D$ and the acceptor $I_A$ ($\eta$ being a quantum yield correction factor), depends on the distance between the two fluorophores according to the equation: $E = 1/(1 + (R/R_{0})^{6})$, where $R$ is the distance between both fluorophores and $R_{0}$ is the characteristic distance (Förster distance) where the efficiency is one-half \citep{lakowicz2006principles}.

The experiments can be carried out either on-surface-immobilized (Fig. \ref{fig:Tech}A) or freely diffusing molecules. The former permits parallelized measurements during a long time (until photobleaching, typically up to tens of seconds) allowing the detection of slow conformational changes. smFRET requires biochemical modifications by attaching fluorophores to the molecule under study, which may interfere with the (unmodified) molecular behavior \citep{ha1999ligand,zhao2009rna}. In addition, in the case of on-surface-immobilized smFRET experiments, the perturbations that the surface may exert on the molecule must be taken into account\citep{schmitz2015strategy}. There are several works where smFRET has been key to determining intermediate or misfolded states in RNA folding \citep{bartley2003exploration, bokinsky2003single,xie2004single,aleman2008exploring,bokinsky2003single}.

\subsection{Nanopore Microscopy} 
\label{subsec:nanopore}
In recent years, nanopore microscopy has shown to be a promising tool to address RNA complexity. Nanopores, either solid-state or biological ones, are extremely sensitive to the sequence and structure of biomolecules. This technique is based on the current established between two electrodes placed in different pools connected by a nanometric hole. The pools are filled with a salt solution and a charged biomolecule is placed in one of them. When a voltage difference is applied, the flow of ions through the nanopore results in an electric current. The flow of ions and the electric field between the electrodes makes the molecules flow through the nanopore. As molecular translation proceeds, the flow of ions is partially (or fully) obstructed, resulting in a reduction of the net current. These ion fluctuations depend on the biomolecule's properties. Usually, the biomolecule needs to rearrange to translocate, making nanopore translocation an important tool to screen different molecular conformations. 

Traditionally, RNA translocation experiments use biological pores such as ion channel proteins (Fig. \ref{fig:Tech}B, top). A widely used model is alpha-hemolysin \citep{butler2007ionic,sultan2019nanopore,meller2006dna} due to its large stability and small pore diameter that confers much sensitivity to detect individual nucleotides. Biological pores generally exhibit lower noise. However, solid-state nanopores (Fig. \ref{fig:Tech}B, bottom) can operate at higher voltages and bandwidths, making it possible to achieve a better signal-to-noise ratio, key to detecting translocation events. Moreover, solid-state nanopores are pore-size and geometry-tunable and more robust than biological ones.

A key advantage of solid-state nanopore microscopy is that it does not require chemically modifying the biomolecule. However, there are some disadvantages such as the limited nanopores reproducibility and the high translocation speed.

\subsection{Optical Tweezers} 
\label{subsec:OT}

Among all single-molecule force spectroscopy techniques, optical tweezers have proved to be one of the most powerful for studying the complexity of nucleic acids, especially in the case of RNA. Laser optical tweezers (LOT) use a focused laser beam to optically trap a transparent microbead attached to one end of an RNA molecule. By attaching the other end to a surface, the RNA can be pulled by displacing the optically trapped bead by moving the laser. LOT can exert forces in the range of tenths to hundreds of piconewtons measuring energies with 0.1kcal/mol accuracy. LOT can monitor force and bead position, detecting conformational transitions with sub-millisecond temporal resolution.

There are different setups to carry out LOT experiments, such as dual-trap (Fig. \ref{fig:Tech}C, left) and single-trap configurations (Fig. \ref{fig:Tech}C, right). In the dual-trap configuration, the molecule is tethered between two trapped beads, while in the single-trap configuration, the molecule is tethered between two beads: one is optically trapped and the other is held by air suction at the tip of a micropipette. Usually, the dual-trap setup is more sensitive because the traps are formed from the same laser and have less drift. The single-trap micropipette configuration is used in many experiments, such as those presented in this review.

Independently on the setup, to avoid bead-bead interactions, the RNA molecule under study is inserted between two molecular linkers or spacers  usually DNA-RNA hybrid handles \citep{wen2007force,manosas2007force,white2011optical,collin2005verification, liphardt2001reversible, onoa2003identifying, martinez2022measurement,wu2014folding}. 
As compared with other single-molecule techniques, the mechanical manipulation of RNAs allows monitoring the unfolding/folding of individual structural domains of large RNAs and their folding pathways. Moreover, RNAs can be pulled to detect short-lived intermediates and misfolded states, which can be modified by the ion concentration (Na$^{+}$,Mg$^{+2}$, etc.) or the temperature. In addition, RNA manipulation with optical tweezers permits measuring energies with 0.1kcal/mol accuracy \citep{severino2019efficient,rissone2022stem,martinez2022measurement}.

\section{RNA Pulling in a Nutshell}
\label{sec:RNApulling}
Several pulling protocols can be implemented to study RNAs with optical tweezers. The most common are force-ramp, hopping, and force-jump experiments. In the following, we show examples of RNA pulling experiments with LOT.

\begin{figure}[h]
\centering
\includegraphics[width=\textwidth]{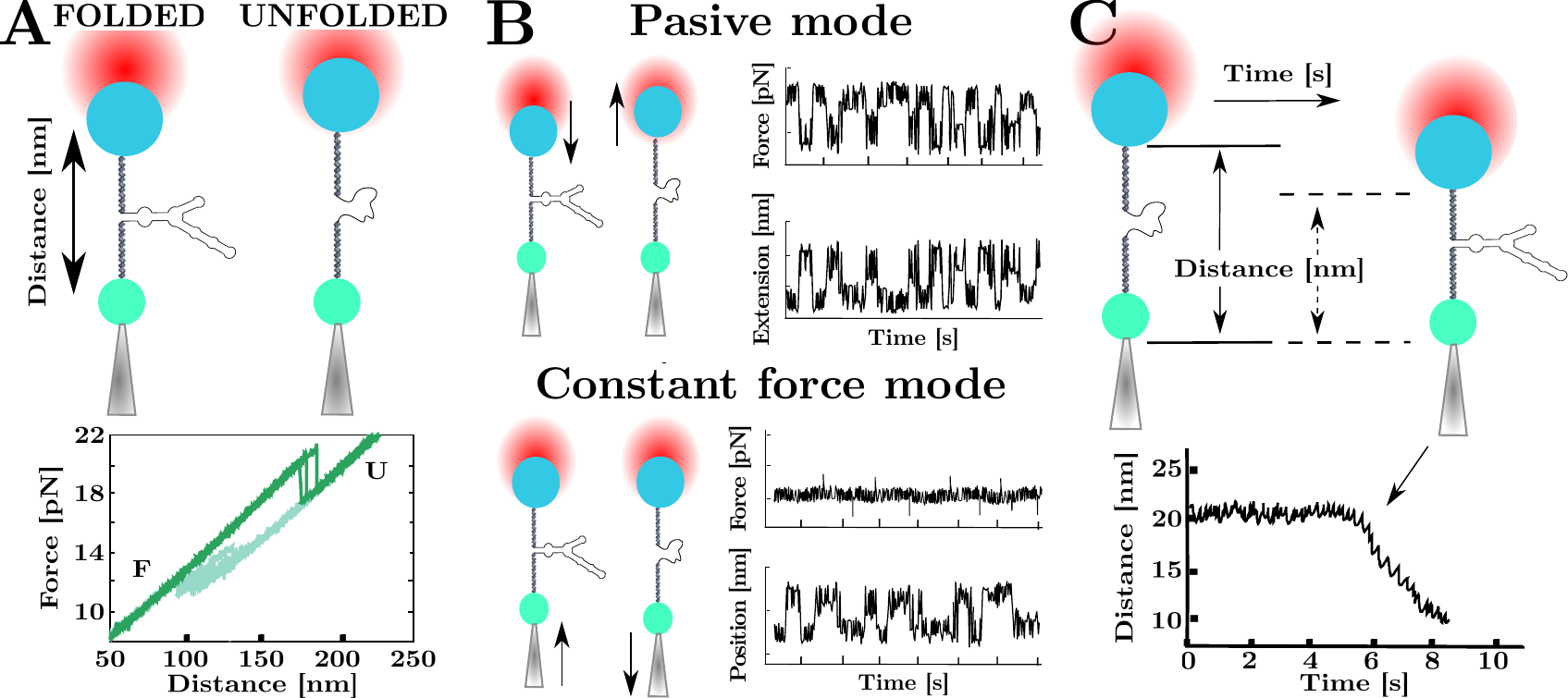}
\caption{\label{fig:Manipulation} Optical tweezers experiments. (\textbf{A}) Force-ramp protocol. The trap-pipette distance is changed at a constant velocity. The typical FDC obtained in force-ramp experiments is shown in the bottom panel. (\textbf{B}) Hopping protocols. Top: PM experiments. The optically trapped bead moves toward or away from the trap center as the molecule unfolds or folds, respectively. The right panels show force and extension variation with time (top and bottom, respectively). Bottom: CFM experiments. The force is kept constant by a feedback protocol compensating for extension changes of the molecule. The right panels show force and position variation with time (top and bottom, respectively). (\textbf{C}) Force jumps protocol. The force or the distance is quickly changed to a different value. The typical distance versus time signal is reported in the bottom panel.} 
\end{figure}

In force-ramp protocols, the optical trap is repeatedly moved back and forth from the micropipette (Fig. \ref{fig:Manipulation}A, top) with the force steadily increasing and decreasing, respectively. Upon increasing the force the RNA hairpin switches from its folded (native) state to the totally unfolded state and \textit{vice-versa} upon releasing the force.
A plot of the applied force versus the distance between the trap and the micropipette gives the force-distance curve (FDC) (Fig. \ref{fig:Manipulation}A, bottom). In small RNA hairpins (typically a few tens of bases) that cooperatively fold and unfold, the rips in the FDC indicate unfolding or folding transitions between the native folded hairpin and the stretched ssRNA conformation. For longer RNA hairpins (a few hundred bases) the unfolding into ssRNA is a sequential process in which groups of base-pairs  open. In this case, the FDC exhibits a sawtooth pattern depending on the molecular sequence (Fig. \ref{fig:FDCs}).
Pulling experiments allows for the characterization of the unfolding/folding forces as well as the thermodynamics and kinetics of RNA molecules (see Sections \ref{sec:energetics} and \ref{sec:dynamics}).

In hopping experiments, either the trap position or the force is kept constant, and the jumps of the molecule from the unfolded (folded) to folded (unfolded) state are monitored over time. There are two kinds of hopping assays: passive mode (PM) and constant force mode (CFM) experiments.
In PM hopping experiments, the trap position is clamped (Fig. \ref{fig:Manipulation}B, top), and the RNA hairpin hops between the unfolded and the folded state with a force jump at every transition. In PM both the force and the molecular extension change over time because the trapped bead relaxes to a new position at every RNA hop, and the force changes accordingly. The PM allows direct monitoring of the molecular transitions in a narrow range of forces close to the coexistence force where the RNA equally populates the native and unfolded states. PM is not suitable for long-time measurements in LOT in the single-trap configuration due to the uncontrolled movements of the pipette (drift effects). 
In contrast, in CFM hopping experiments, the force is kept constant with a feedback loop (Fig. \ref{fig:Manipulation}B bottom), and the RNA extension is recorded as a function of the time. Here, the force is fixed at a value near the RNA coexistence force, allowing for measuring the transition between its folded/unfolded states and the lifetime of each state at the studied force. 

The force-jump protocol is helpful in characterizing RNA irreversible processes. In this case (Fig. \ref{fig:Manipulation}C), the force or the trap position is quickly changed to a new preset value and kept constant. The RNA lifetime is measured until a conformational transition is observed. Force-jump is an irreversible pulling protocol, requiring multiple experiments at different present force values to derive the unfolding and folding kinetics.

\subsection{Brief History of RNA Pulling Experiments}
\label{subsec:history}

\begin{figure}[t]
\centering
\includegraphics[width=\textwidth]{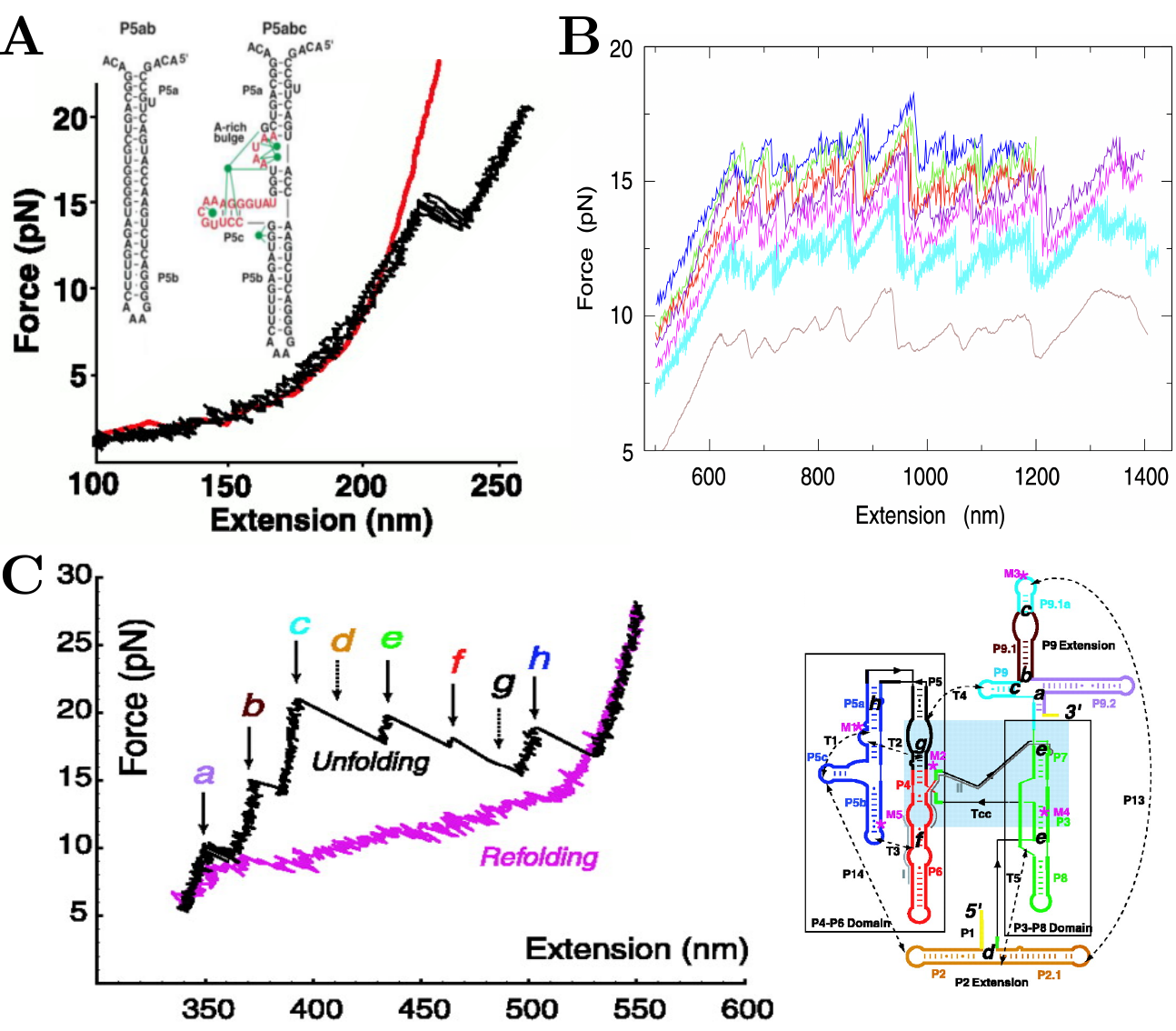}
\caption{\label{fig:History1} Different results obtained in pulling experiments. (\textbf{A}) P5ab RNA force-extension curves in 10 mM Mg$^{2}$ (adapted from \citep{liphardt2001reversible}). Structure of P5ab and P5abc RNAs (inset). (\textbf{B}) 16S ribosomal FDCs, the colors represent  successive pulling for the same molecule. Figure adapted from \citep{harlepp2003probing}. (\textbf{C}) Unfolding (black) and folding (pink) FDCs of the \textit{T. thermophila} ribozyme (left panel).  The letters indicate different kinetics barriers. The right panel shows the secondary structure. Figure adapted from \citep{onoa2003identifying}.} 
\end{figure}

Since the beginning of the 21$^{\rm st}$ century, LOT experiments have been used in many works studying the structure and energetics of RNAs and their native, misfolded, and short-lived intermediate states, and the salt dependency of the RNA conformation. 
In \citep{liphardt2001reversible}, it has been shown for the first time that by mechanically unzipping RNA hairpins, it is possible to derive its folding free energy (Fig. \ref{fig:History1}A). They studied three RNAs: a small hairpin (P5ab), a molecule with a three-helix junction (P5abc$\Delta$A), and a more complex molecule (P5abc), which in presence of Mg$^{+2}$ ions folded into a stable tertiary structure. In particular, they found that P5abc folded into a stable tertiary structure through a short-lifetime intermediate.

Mechanical pulling experiments have also been carried out on longer RNAs: 
in \citep{harlepp2003probing}, it has been studied the 1540 nucleotides 16S ribosomal RNA from {\it E. coli} (Fig. \ref{fig:History1}B). This large RNA exhibits a surprisingly well-structured and reproducible unfolding pathway under mechanical stretching. 
Similar results have been found in \citep{onoa2003identifying} by pulling the L-21 derivative of the \textit{Tetrahymena thermophila} ribozyme (a 390 nucleotides RNA). This molecule featured a complex secondary structure with multiple unfolding intermediate states (Fig. \ref{fig:History1}C). To identify the different RNA structures, they measured the number of released base pairs along the FDCs of progressively large fragments of the RNA molecule. At the same time, they used mutants and anti-sense oligonucleotides to characterize these RNA structures further and measure their free energy of formation.

\begin{figure}[t]
\centering
\includegraphics[width=\textwidth]{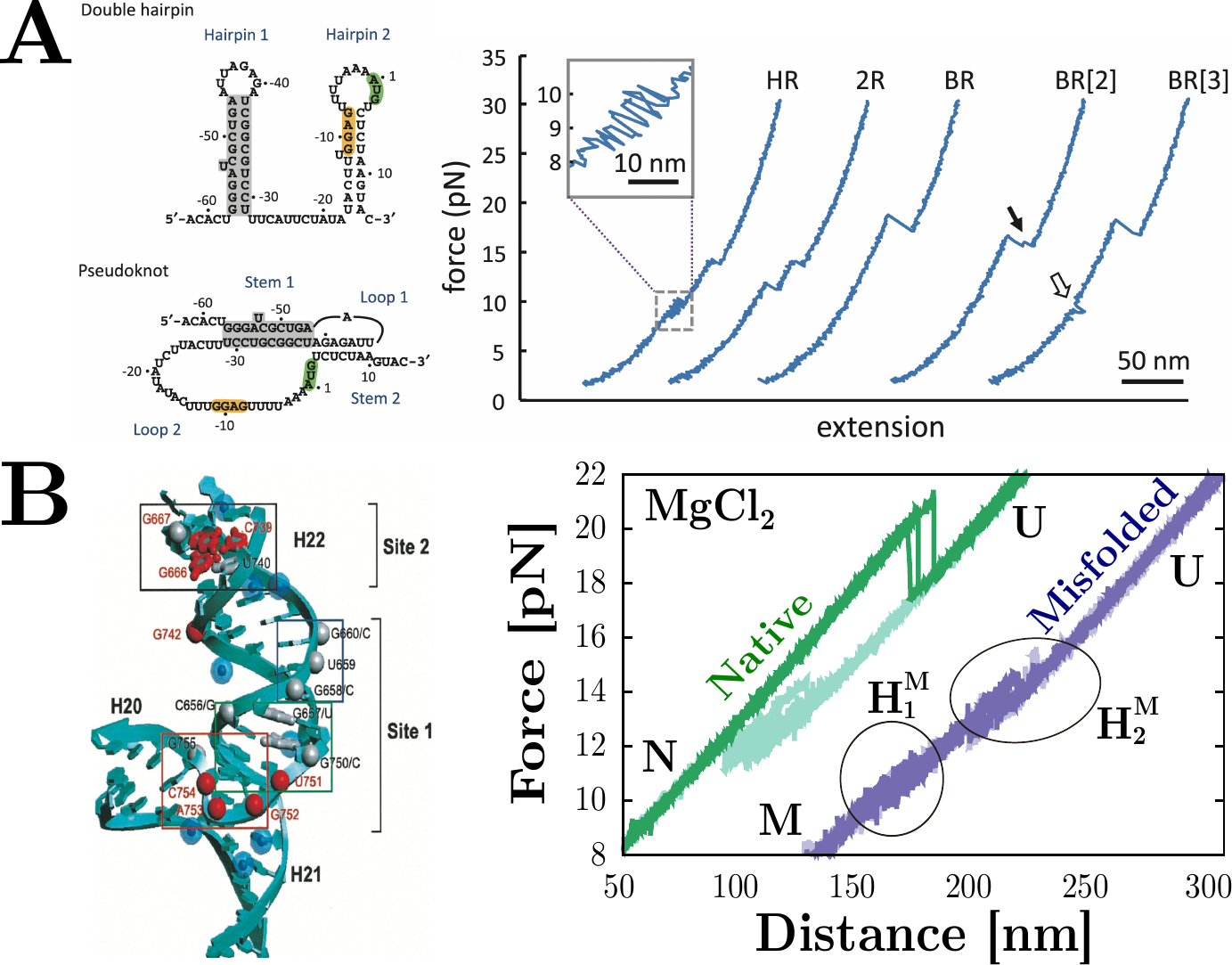}
\caption{\label{fig:History2} Different results obtained in pulling experiments. (\textbf{A}) Left: the two structures proposed for rpsO RNA: double hairpin (top) and pseudoknot (bottom). Right: FDCs of the double hairpin (HR), pseudoknot (BR), and mixed transitions (2R). Figure adapted from \citep{wu2014folding}. (\textbf{B}) Left: The highly conserved crystal structure of the RNA three-way junction molecule from \textit{T. thermophilus} \citep{serganov2001role}. Right: FDCs for the native and misfolded structures in 10 mM Mg$^{+2}$. Figure adapted from \citep{martinez2022measurement}.} 
\end{figure}

Another example of a biologically relevant RNA is the study of the operator rpsO of the RNA gene coding for the S15 subunit of 30S ribosomal protein from {\it E. coli} \citep{wu2014folding}. This mRNA can fold into two spontaneously interchangeable structures: a double hairpin and a pseudoknot (Fig. \ref{fig:History2}A, left). Their work showed that the conversion from the double hairpin to the pseudoknot is stabilized by the interaction between the two hairpins (Fig. \ref{fig:History2}A, right). Sequence mutations can modulate the interaction between the two hairpins.

Finally, in a more recent work \citep{martinez2022measurement} RNA pulling experiments have been carried out to measure the non-specific and specific binding energies of magnesium to RNA. The RNA three-way junction (3WJ) containing the minimal binding site to protein S15 of the ribosomal RNA from {\it E. coli} was studied in monovalent and divalent salt conditions.  This molecule can fold into a native structure that contains the 3WJ motif with specific Mg$^{+2}$ binding sites or into a misfolded structure with a double hairpin structure where the 3WJ and the binding sites have been disrupted (Fig. \ref{fig:History2}B). By comparing the free energy of formation of these two structures with and without magnesium, it has been possible to determine the specific and non-specific energy contributions of Mg$^{+2}$ binding to the RNA.

%
\section{RNA Energetics at $0.1$ kcal/mol Accuracy}
\label{sec:energetics}

\begin{figure}[t!]
\centering
\includegraphics[width=\textwidth]{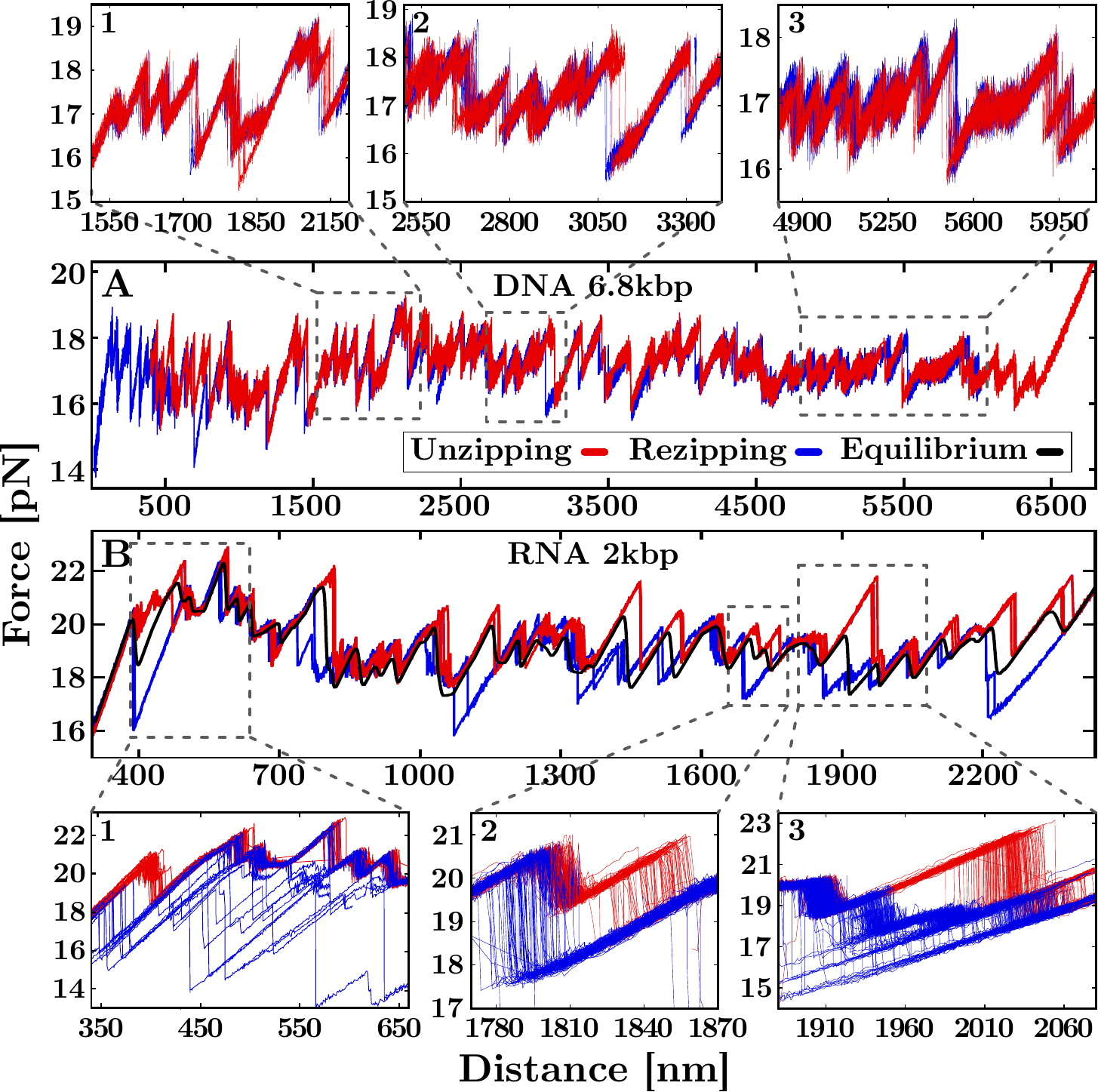}
\caption{\label{fig:FDCs} Unzipping (red) and rezipping (blue) FDCs measured by pulling a 6.8kbp DNA (\textit{A}) and a 2kbp RNA (\textit{B}) hairpins in a 1M NaCl buffer at $25^{\circ}$~C. While DNA unzipping is a reversible process, RNA unzipping shows many off-pathway out-of-equilibrium states. The resulting hysteresis requires the computation of an equilibrium FDC (black) to measure the NNBPs' free energies.} 
\end{figure}

The characterization of RNA thermodynamics is fundamental to understanding the promiscuity of behaviors observed in RNA, from the multiplicity of native structures \citep{gralla1974biological} to misfolding \citep{alemany2012experimental}.
Only RNA exhibits such behavior despite DNA and RNA forming double-stranded helices. DNA unzipping is a fully reversible process (Fig. \ref{fig:FDCs}A) over a broad range of salt conditions and loading rates \citep{huguet2010single,bizarro2012non}. Instead, RNA unzipping presents strong irreversibility between the unfolding and refolding FDCs in the same experimental conditions (Fig. \ref{fig:FDCs}B). In this case, transient off-pathway misfolded structures appear during the unzipping–rezipping process, slowing down the hybridization reaction \citep{liphardt2001reversible,chen2000rna,woodson2010compact}. Characterizing these off-pathway structures competing with the native stem is a challenging problem \citep{rissone2022stem}.

%

\subsection{The RNA Free Energy of Formation}
\label{subsec:NNenergies}

The hybridization of the double-stranded helix of nucleic acids (NA) is governed by the Watson-Crick pairing rules between nucleotides (adenine, guanine, cytosine, thymine or uracil) from opposite NA strands \citep{saenger1984principles}. These relations account for purine–pyrimidine bonding so that adenine can only bond to thymine (or uracil in the RNA case) and guanine to cytosine.
As discussed before, the double helix is not only stabilized by base-pairing but also by the stacking interactions between adjacent nucleotides. To account for these effects, the energetics of NA formation is usually described using the nearest-neighbor (NN) model \citep{devoe1962stability,breslauer1986predicting}. According to this model, the base-pairing energy of two complementary bases depends on the base itself and the first neighbor located in the same strand along the $5^{\prime} \rightarrow 3^{\prime}$ direction. This gives 16 different possible combinations of nearest-neighbour base-pairs (NNBPs): out of these, 6 are degenerate, and 2 can be expressed as a linear combination of the others (circular symmetry) \citep{huguet2017derivation}, leaving with only 8 independent parameters. The ten RNA NNBP are denoted as $\rm XY/\overline{X}\,\overline{Y}$ (x-label in Fig. \ref{fig:kinetics}A) where $\rm X,Y=A,C,G,U$, and $\rm \overline{X}(\overline{Y})$ is the complementary base of $\rm X(Y)$ and $\rm {XY/\overline{X}\,\overline{Y}}$ is the NNBP resulting from hybridizing dinucleotides $\rm 5^{\prime}-XY-3^{\prime}$ and $\rm 5^{\prime}-\overline{Y}\,\overline{X}-3^{\prime}$. The energies of $\rm {XY/\overline{X}\,\overline{Y}}$ and $\rm \overline{Y}\,\overline{X}/YX$ are equal due to complementary strand symmetry. For example, the RNA sequence $\rm 5^{\prime}-CUUAGC-3^{\prime}$ forms a duplex with its complementary strand, $\rm 5^{\prime}-GCUAAG-3^{\prime}$. According to the NN model the energy of hybridization of such a sequence equals $\rm \Delta g_{CU/GA} + \Delta g_{UU/AA} + \Delta g_{UA/AU} + \Delta g_{AG/UC} + \Delta g_{GC/CG}$ with $\rm \Delta g_{CU/GA}=\Delta g_{AG/UC}$ due to complementary strand symmetry.
The derivation of the 10 (8 if circular symmetry is applied) NNBPs free-energies has been carried out by unzipping a 2.2kbp and 6.8kbp DNA and a 2kbp RNA hairpins at different salt conditions \citep{huguet2010single,huguet2017derivation,rissone2022stem}. 

In thermodynamics, the free-energy difference equals the mechanical work in a reversible process that requires the system to evolve along a sequence of equilibrium states. This is only applicable to the DNA case where the unzipping and rezipping FDCs do not show hysteresis (Fig. \ref{fig:FDCs}A). On the contrary, RNA unzipping is an out-of-equilibrium process in our experimental timescales: a large hysteresis is observed between unzipping and rezipping FDCs due to the formation of multiple long-lived (off-pathway) states (Fig. \ref{fig:FDCs}B). 
An equilibrium FDC (black line) had to be computed from the RNA unzipping/rezipping experimental data. This has been achieved by developing a statistical method based on the (extended) fluctuation relations \citep{bennett1976efficient,jarzynski1997nonequilibrium,shirts2003equilibrium,junier2009recovery} that allowed for the computation of the equilibrium free energy during the isothermal unzipping process (see Supp. Info in \citep{rissone2022stem}).
Finally, a Monte Carlo optimization algorithm has been tailored to relate the experimental data with the numerical FDC prediction, ultimately permitting measuring the NNBP energies in DNA and RNA \citep{huguet2010single,huguet2017derivation,rissone2022nucleic}. 
The experimentally derived values for RNA in sodium and magnesium are shown in Fig. \ref{fig:energetics}A along with values reported in the literature (the Mfold set) \citep{freier1986improved,mathews1999expanded,walter1994coaxial,xia1998thermodynamic,zuker2003mfold}. As the latter are only available at 1M NaCl, the comparison required applying a correction to the measured NNBP energies.

\subsection{Salt Dependency of the Hybridization Free Energy}
\label{subsec:salt_rule}

The effect of a monovalent salt concentration, $[\rm Mon+]$ in molar units, on the hairpin free energy of formation is described by the relation
\begin{equation}
    \label{eq:salt_corr}
    \Delta g_{0,i} [{\rm Mon+}]  =  \Delta g_{0,i} [{\rm 1 M}] - m \log{[{\rm Mon+}]} \, ,
\end{equation}
where $\Delta g_{0,i} [\rm 1 M]$ is the free energy of formation of motif $i$ at 1M of monovalent salt at zero force and $m=0.10\pm0.01$~kcal/mol is the RNA NNBP-homogeneous monovalent salt correction \citep{bizarro2012non}.
However, Eq.\eqref{eq:salt_corr} only holds for monovalent ions and its extension to divalent ions requires to account for the effect of the divalent salt on the stabilization of the double-helix.

\begin{figure}[t]
\centering
\includegraphics[width=0.95\textwidth]{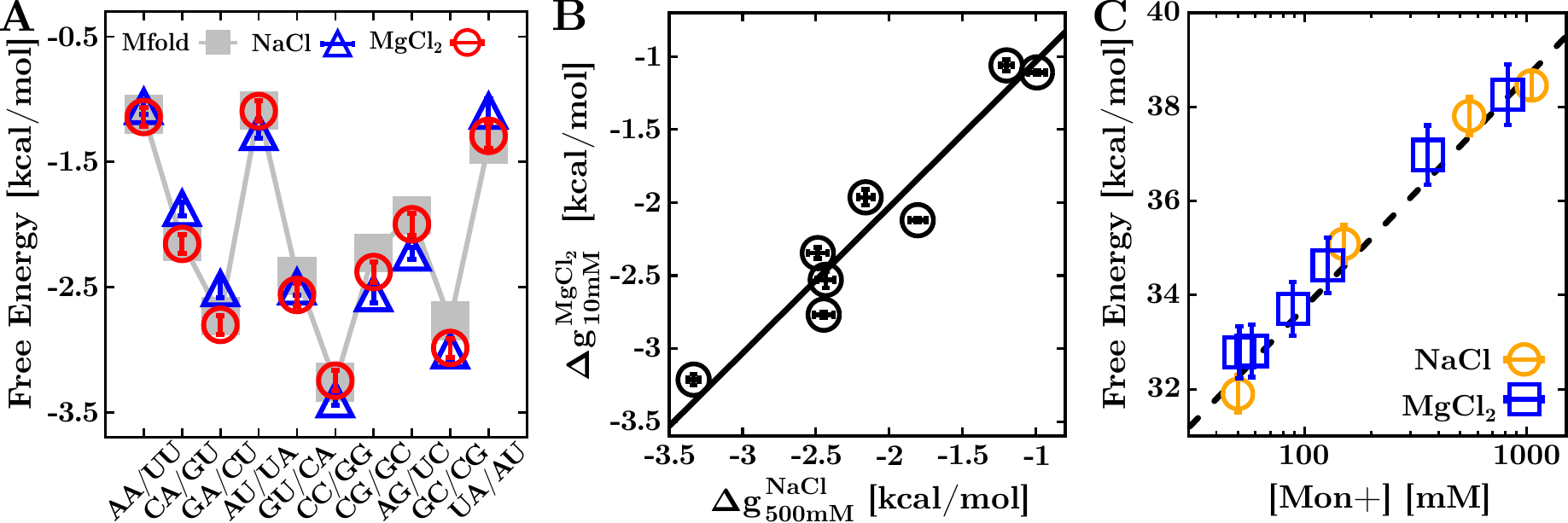}
\caption{\label{fig:energetics} RNA energetics. ($\mathbf{A}$) NNBPs free-energies measured by unzipping a 2kbp RNA hairpin at 500mM NaCl (triangles) and 10mM MgCl$_2$ (circles) \citep{rissone2022stem}. Notice that the values are scaled to 1M of equivalent sodium concentration in order to compare with the literature (grey squares). ($\mathbf{B}$) Experimental validation of the 100:1 rule of thumb. A fit to data (solid black line) gives $77(\pm 49)$:1 \citep{rissone2022stem}. ($\mathbf{C}$) Hairpin (total) free-energy of formation measured in RNA unzipping experiments of a 20bp hairpin \citep{bizarro2012non} in sodium (blue squares) at 50mM, 150mM, 550mM, 1050mM and magnesium (orange circles) at 0.01mM, 0.10mM, 0.50mM, 1mM, 4mM, 10mM. A fit to data (dashed line) shows the logarithmic dependence of the salt correction. All results in magnesium are reported in sodium equivalents according to the 100:1 rule.} 
\end{figure}

RNAs are highly charged polyanions whose stability strongly depends on solvent ionic conditions. The ability of divalent ions, such as $\rm Mg^{2+}$, to stabilize RNA structures at much lower concentrations than monovalent ions is known since the 70\emph{s} \citep{cole1972conformational}. This effect is usually quantified by the so-called 100:1 rule of thumb which states that the concentration of divalent salt equals 100-fold that of monovalent salt.
This phenomenological rule has been experimentally tested by unzipping the 2kbp RNA hairpin at 500 mM NaCl and 10 mM MgCl$_2$ \citep{rissone2022stem}. By plotting the RNA NNBP energy values in $\rm Na^{+}$ versus those in  $\rm Mg^{2+}$ (Fig. \ref{fig:energetics}B), we demonstrate that the divalent salt concentration is equal to $77\pm49$ the one of monovalent salt, which is compatible with the phenomenological rule.

This result has also been validated by measuring the free energy of formation of short RNA duplexes unzipped in a broad range of monovalent and divalent salt concentrations \citep{bizarro2012non}. By plotting these results versus the salt concentration, the data can be collapsed into a single master curve by multiplying the $[{\rm Div}\, 2+]$ by a factor $\sim 80$ (Fig. \ref{fig:energetics}C).
The agreement between the results proves that the RNA hybridization free energy of the hairpin, i.e. the energy of the hairpin native conformation, satisfies (within errors) the 100:1 rule of thumb. Finally, a fit to data (dashed grey line) proved the validity of the logarithmic salt correction to the formation energy in Eq.\eqref{eq:salt_corr}.

%

%
\section{RNA Folding Kinetics}
\label{sec:dynamics}
\begin{figure}[t]
\centering
\includegraphics[width=\textwidth]{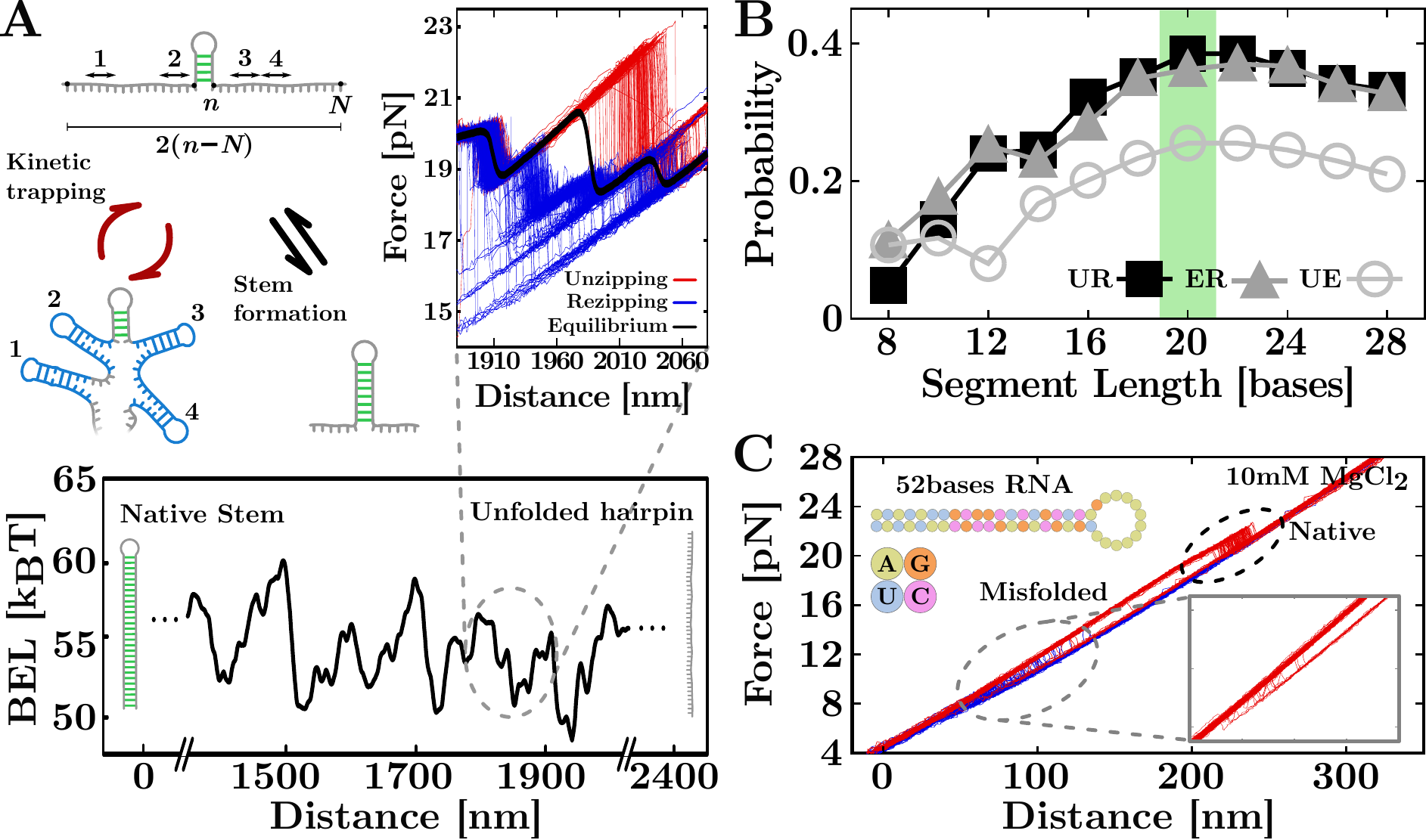}
\caption{\label{fig:kinetics} RNA unzipping and rezipping kinetics. ($\mathbf{A}$) Stem-loops BEL of a 2kbp RNA hairpin. The stem-loops formation along the ssRNA drives the folding process by stabilizing off-pathway states that slow down the stem hybridization (top-left). The BEL correlates with the hysteresis measured along the FDC (bottom and top-right inset). Panel adapted from \citep{rissone2022nucleic}. ($\mathbf{B}$) Correlation of the forming stem-loops with the UR, ER, and UE hysteresis (see text) as a function of stem-loops length, $L$. ($\mathbf{C}$) Unzipping of a short (52 bases) RNA hairpin at 10 mM MgCl$_2$. The presence of divalent ions causes the molecule to misfold (inset).}
\end{figure}

The typical unzipping patterns of DNA and RNA hairpins are very different, as shown in Fig. \ref{fig:FDCs}A and Fig. \ref{fig:FDCs}B. To characterize the strong irreversibility observed in RNA, it has been hypothesized that stem-loop structures form along the two unpaired RNA strands during the unzipping (rezipping) process.
As the unzipping reaction progresses, forming such structures close to the junction (that separates the native stem from the unpaired ssRNA) slows down hairpin hybridization. Consequently, the system gets trapped into off-pathway metastable conformations generating the observed hysteresis (Fig. \ref{fig:kinetics}A, top-left).

This scenario can be modeled by introducing the barrier energy landscape (BEL) \citep{rissone2022stem,rissone2022nucleic} that includes all combinations of a number $k$ of stem-loops ($k=1,2,\dots K$, with $K$ a maximum total number) that can form in the two unpaired RNA strands. Let $n$ be the number of hybridized base pairs in a hairpin with a total number of N base pairs (Fig. \ref{fig:kinetics}A, top-left). Each unpaired strand contains N-n bases, and the two unpaired strands taken together contain 2(N-n) bases. Therefore, we can consider that stem loops can form in a single unpaired strand of length 2(N-n). For a given number $k$ of stem-loops of size $L$, we will consider them as randomly distributed along the 2(N-n) bases strand (Fig.A). For a given applied force $f$, the BEL is defined as 
\begin{equation}
    \label{eq:BEL}
    \Delta G_L(f) = -k_{\rm B}T \log \sum_{k=0}^{K} \exp \left(-\frac{\Delta g_{L}(k,f)}{k_{\rm B}T} \right) \, ,
\end{equation}
where $\Delta g_{L}(k,f)$ is the total free-energy contribution of $k\ge 0$ stem-loops along the unpaired strand and $K = \lfloor 2(N-n)/L \rfloor$ is the maximum number of stem-loops \citep{rissone2022nucleic}. Notice that for $k=0$ no stem loops are formed. In principle, stem loops cannot overlap because only one structure can be formed with the same bases. For sake of simplicity, we will consider the approximated case where stem loops can overlap.

The loop-BEL computed in Eq.\eqref{eq:BEL} correlates with the amount of observed hysteresis (Fig. \ref{fig:kinetics}A, bottom and top-right inset), measured as the area between unzipping-rezipping (UR), unzipping-equilibrium (UE) and equilibrium-rezipping (ER) curves \citep{rissone2022stem}. 
In particular, the study of this correlation as a function of the stem-loops size, $L$, shows that the stability of stem-loops increases with $L$ reaching a maximum for $L\sim 20$ bases and gently decaying for larger sizes (Fig. \ref{fig:kinetics}B).

Stem-loops formation is a kinetic process that competes with native hybridization explaining the different behaviors observed in DNA and RNA unzipping. A main source of irreversibility in RNA is the higher stability of RNA stacking. In fact, the average  NNBP RNA free-energy is $\langle \Delta g_0 \rangle^{{\rm RNA}} \approx -2.2$~kcal/mol while for DNA $\langle \Delta g_0 \rangle^{{\rm DNA}} \approx -1.7$~kcal/mol.
This $\approx 0.5$~kcal/mol difference could be sufficient for the RNA hairpin to slow down hybridization without the need for stem loops forming along the unpaired strands making the loop-BEL unnecessary.  
However, there is mounting evidence that RNAs can form a multiplicity of structures as compared to DNA. In particular, in \citep{rissone2022stem} it was found that a 52 bases native hairpin can also form alternative misfolded structures not predicted by secondary-structure RNA numerical models.  
By competing with the native pathway of the 2.2kb RNA hairpin, the stem loops slow down the hybridization reaction stabilizing the on-pathway intermediates in the FDC leading to the observed hysteresis.

Moreover, the observed irreversibility is enhanced in presence of magnesium \citep{rissone2022stem}. The metal ions induce higher flexibility to the RNA chain and higher stability to the RNA helices due to the coordination effect of the two positive charges and a reduction in the backbone's charge repulsion. An essential consequence of the reduced charge repulsion is that it allows more frequent close encounters between the different RNA segments, facilitating the formation of tertiary contacts  \citep{tan2009predicting}.
Several studies pointed out that $\rm Mg^{2+}$ ions are more efficient than $\rm Na^{+}$ in stabilizing RNA tertiary folds \citep{chu2007evaluation,chen2008rna,draper2008rna,walter2008non}.
The same phenomenon is not observed for DNA, where salt concentrations as high as 10 mM $\rm Mg^{2+}$ and 1 M $\rm Na^{+}$ (equivalent concentrations as per the salt rule) induce similar foldings \citep{tan2007rna}. 
On the contrary, unzipping experiments at 10 mM $\rm Mg^{2+}$ on the previously mentioned 52 bases RNA hairpin showed that magnesium induces misfolding (Fig. \ref{fig:kinetics}C), whereas only the native conformation is present at the equivalent concentration of 1 M $\rm Na^{+}$ \citep{rissone2022stem}. In general, it has been observed that in mixed conditions of monovalent and divalent salt, low concentrations of the latter are sufficient to stabilize RNA tertiary structures in presence of specific binding sites  \citep{heilman2001role}.

\section{Future Perspectives}
\label{sec:future}
Force spectroscopy is an exquisite tool to probe chemical interactions in biomolecules. The finely tuned balance between hydrogen bonding and stacking energies in nucleic acids makes them extremely sensitive to environmental changes, sequence mutations, and chemical modifications \citep{song2017chemical}. Single-molecule fluorescence and nanopore microscopy are complementary tools that permit the detection and monitor conformational transitions at zero force. In contrast, force spectroscopy cannot probe molecular states at zero force because the end-to-end molecular extension cannot be detected.  Therefore the combination of force with fluorescence (optical fleezers) \citep{whitley2017high} and nanopores (optical trap nanopore) \citep{keyser2006optical, trepagnier2007controlling, yuan2020controlling} offers exciting prospects for the future. 

Force spectroscopy permits the direct estimation of free energy differences at room temperature by direct work measurements, $W=\Delta G$. A new direction of expansion is now possible using temperature-controlled optical tweezers \citep{mao2005temperature, mahamdeh2009optical, de2015temperature}. Controlling temperature is often tricky, especially when the heating region is large compared to the typical micrometer-sized dimensions of the experimental trapping area. In this case, thermal expansion and convection effects lead to uncontrolled drift effects. In a recent setup \citep{de2015temperature}, a heating laser can be switched on and off to controllably heat up the experimental trapping region without convection and drift effects. The instrument also operates at low temperatures by using an icebox kept at water-freezing temperatures (1-4$^{\circ}$C). This temperature-jump optical trap has been recently used to derive folding enthalpies, and entropies of DNA \citep{de2015temperature} and proteins \citep{rico2022molten}. The instrument permits measuring heat capacity changes, cold denaturation, and other previously inaccessible phenomena. Fascinating is the study of RNA at very low temperatures where thermal fluctuations are reduced, folding kinetics slowed down, and details of the molecular interactions intensified. By lowering the temperature, monitoring kinetics provides a natural microscope to amplify the finest energetic features driving RNA folding.

While DNA is often referred to as life's molecule \citep{frank1993unraveling}, RNA is the dark matter of the genome \citep{darnell2011rna, mattick2022rna}, underlining how much we still do not know about this fascinating molecule. RNA presents such remarkable features that biophysical models must be refined to understand its many behaviors. We need physical models based on molecular energy landscapes to unravel the folding kinetics for RNA folding. Concepts borrowed from physics such as rugged free energy landscapes \citep{kirkpatrick2015colloquium}, and molecular replica symmetry breaking \citep{ritort2022molecular}. Ideas from soft and condensed matter physics may come into play shortly to explain why RNA is so unique, in stark difference from the stable DNA counterpart. The critical question is understanding under which conditions DNA might behave as RNA. DNA lacks the reactive $2^{\prime}$--OH from ribose in RNA, impairing its catalytic activity. Yet, DNA might fold in some conditions as RNA does. Foreseeable experiments in the future are the study of folding kinetics of ssRNAs and differences with ssDNA \citep{viader2021cooperativity, rissone2022nucleic}. Other unexplored areas are awaiting discovery, such as RNAs with chemical modifications, RNAs at low temperatures, RNAs in crowded environments, etc. From custom-designed RNAs to biological RNAs (coding and non-coding), our knowledge of this fantastic molecule has never stopped growing and will be so for decades.

\begin{acknowledgement} 
P.R. was supported by the Angelo Della Riccia foundation. I. P. and F.R. were supported by Spanish Research Council Grant PID2019-111148GB-I00 and the Institució Catalana de Recerca i Estudis Avançats (F. R., Academia Prize 2018).
\end{acknowledgement}

\label{sec:refs}
\bibliographystyle{spbasic}
\bibliography{BibliographyEDITED}

\end{document}